\documentclass[
reprint,
superscriptaddress,
amsmath,amssymb,
aps,
]{revtex4-2}

\usepackage{color}
\usepackage{graphicx}
\usepackage{dcolumn}
\usepackage{bm,braket}

\begin {document}

\title{
    Sample-specific rectification-like response in a boundary-driven exclusion process
    }

\author{Issei Sakai}
\email{isseisakai5@gmail.com}
\affiliation{%
    Department of Physics and Astronomy, Tokyo University of Science, Noda, Chiba 278-8510, Japan
}%

\author{Takuma Akimoto}
\email{takuma@rs.tus.ac.jp}
\affiliation{%
    Department of Physics and Astronomy, Tokyo University of Science, Noda, Chiba 278-8510, Japan
}%



\date{\today}

\begin{abstract}
    We investigate the current response of a boundary-driven symmetric exclusion process with quenched site disorder.
    Hard-core particles hop symmetrically on a one-dimensional lattice with site-dependent rates and are injected and removed at the boundaries by two reservoirs of different densities.
    We approximate the steady-state density profile using a Galerkin projection at linear order and a mean-field closure at higher orders, and thereby obtain the current as a nonlinear function of the reservoir density difference.
    At linear order in the reservoir density difference, the current-response coefficient depends on the mean reservoir density $\rho$, in contrast to the homogeneous case.
    Through the linear-response relation, this dependence leads to an equilibrium current-fluctuation coefficient that is asymmetric under $\rho\rightarrow 1-\rho$.
    Beyond linear response, nonzero even-order current contributions break the antisymmetry of the current under reversal of the reservoir density difference, producing rectification-like behavior in individual disorder realizations.
    We further show that spatial-reflection symmetry of the equilibrium density profile rules out such behavior, so broken spatial-reflection symmetry of the profile is a necessary condition for rectification-like behavior.
    Within the present approximation, we further find that, for continuously distributed site disorder, rectification-like behavior occurs arbitrarily close to equilibrium for almost every disorder realization.
    At the ensemble level, however, the disorder-averaged current remains antisymmetric because the disorder ensemble is invariant under spatial reflection.
    These results provide a mechanism for rectification-like transport arising from sample-specific spatial heterogeneity rather than from an explicitly imposed directional asymmetry.
\end{abstract}

\maketitle


\section{Introduction}
Single-file transport is a fundamental transport phenomenon observed in narrow channels, where hard-core particles cannot pass each other.
While the mean-squared displacement of a single Brownian particle grows linearly with time, that of a tracer particle in single-file transport grows sublinearly, specifically as $\Braket{\delta x^2(t)}\propto t^{1/2}$ \cite{Harris_1965}.
This anomalous transport behavior has been observed experimentally in various confined systems, including colloidal particles in one-dimensional channels \cite{doi:10.1126/science.287.5453.625,PhysRevLett.93.026001}, molecular diffusion in zeolites \cite{doi:10.1126/science.272.5262.702,PhysRevLett.76.2762}, and transport through nanopores \cite{PhysRevLett.89.064503,Das:2010aa}.
These observations motivate theoretical studies of how excluded-volume interactions affect collective transport in confined systems.

The symmetric exclusion process (SEP) provides a minimal model for describing single-file transport \cite{SPITZER1970246,Derrida:2007ab,SciPostPhysLectNotes.106}.
In the SEP, particles perform continuous-time random walks on a one-dimensional lattice subject to hard-core exclusion.
When the system is coupled to particle reservoirs, the steady-state current directly characterizes the resulting boundary-driven transport. In a homogeneous open SEP, the steady-state mean current is proportional to the density difference between the reservoirs and is independent of the mean reservoir density~\cite{Derrida:2004aa,PhysRevLett.92.180601}.
The statistical properties of the integrated  current have also been studied extensively, and exact results are available for periodic, open, semi-infinite, and infinite geometries~\cite{PhysRevE.78.021122,Derrida:2004aa,PhysRevLett.92.180601,Saha:2023aa,PhysRevLett.133.117102,Derrida:2009ab,Derrida:2009aa,PhysRevLett.129.040601}.
These results provide a detailed understanding of current statistics and transport in homogeneous exclusion processes.

Most exact results for exclusion processes have been obtained for spatially homogeneous hopping dynamics.
In realistic confined systems, however, transport often occurs in spatially heterogeneous environments arising from structural disorder or disordered energy landscapes.
Such heterogeneity can significantly affect transport properties, leading, for example, to anomalous diffusion~\cite{Bouchaud:1990aa,Metzler:2000aa,Hofling:2013aa} and rectification~\cite{Bhattacharya:2011aa,Manara:2015aa,Zhou:2020aa}.
At the single-particle level, transport in heterogeneous environments has been studied extensively~\cite{10.1039/c4cp03465a,PhysRevLett.117.180602,*PhysRevE.97.052143,PhysRevE.101.042133,PhysRevLett.133.037101}.
By contrast, much less is understood about how heterogeneity affects transport in interacting many-particle systems.
Heterogeneous exclusion processes provide a natural framework for investigating how spatial disorder modifies collective transport in the presence of excluded-volume interactions.

Disordered exclusion processes can be driven out of equilibrium either by applying an external field or by coupling the system to particle reservoirs with different densities.
Under external-field driving, spatial heterogeneity can induce phase separation in the steady-state density profile and suppress the current~\cite{PhysRevLett.78.3039,PhysRevE.58.1911,PhysRevE.70.016108,Bahadoran:2015aa,PhysRevE.107.L052103,PhysRevE.107.054131}.
In our previous work~\cite{21vm-n8gp}, we further showed that site and bond disorder produce qualitatively different responses.
Here, site disorder refers to hopping rates determined by the departure site, whereas bond disorder refers to rates determined by the traversed bond.
A site-disordered system exhibits rectification-like behavior: reversing the external field reverses the current direction but generally changes its magnitude.
This behavior is absent in the corresponding bond-disordered system.

Boundary-driven transport has been studied for bond-disordered exclusion processes, for which the steady-state current remains proportional to the density difference between the reservoirs, as in the homogeneous SEP, although its magnitude is reduced by disorder~\cite{PhysRevLett.80.85}.
Thus, previous studies have examined exclusion processes with spatially varying hopping rates under external-field driving, primarily in particle-number-conserving settings, and boundary-driven systems with bond disorder.
By contrast, the response of a symmetric exclusion process with quenched site disorder to unequal reservoir densities has received comparatively little attention.

In this paper, we investigate how quenched site disorder modifies the current response to a reservoir density difference.
We identify two distinct effects of site disorder on the boundary-driven current response.
First, even at linear order, the current-response coefficient depends on the mean reservoir density $\rho$, in contrast to the homogeneous and bond-disordered cases.
Through the linear-response relation, this dependence also causes the equilibrium current-fluctuation coefficient to be asymmetric under $\rho\rightarrow 1-\rho$.
Second, beyond linear order in the reservoir density difference, even-order contributions can arise, so the current need not be antisymmetric under reversal of the density difference.
This produces rectification-like behavior in individual disorder realizations.
We further establish an exact structural condition for this behavior: spatial-reflection symmetry of the equilibrium density profile guarantees current antisymmetry, so broken reflection symmetry is necessary for rectification-like behavior.
Within the present approximation, the second-order current coefficient is nonzero for almost every realization of continuously distributed site disorder, implying rectification-like behavior arbitrarily close to equilibrium.
At the ensemble level, however, reflection invariance of the disorder distribution restores exact current antisymmetry.
We also examine the distribution and finite-size behavior of the rectification measure and find that its disorder-averaged magnitude decreases with system size over the range studied.

The remainder of this paper is organized as follows.
In Sec.~\ref{sec: model}, we formulate the SEP on a quenched random energy landscape under open boundary conditions and define the steady-state current.
In Sec.~\ref{sec: density profile}, we derive the equilibrium distribution and construct approximations for linear and nonlinear responses of the density profile, which are compared with numerical simulations.
In Sec.~\ref{sec: current}, we derive the linear current-response and current-fluctuation coefficients, obtain the nonlinear steady-state current, establish a necessary condition for rectification-like behavior, and examine its typicality for continuously distributed site disorder.
In Sec.~\ref{sec: disorder ensemble}, we derive the exact disorder-ensemble symmetries and examine the distribution and finite-size behavior of the rectification measure.
In Sec.~\ref{sec: discussion}, we discuss the structural condition required for the rectification-like response and its generality.
Section \ref{sec: conclusion} summarizes our conclusions.

\section{Model}\label{sec: model}
\begin{figure}
    \centering
    \includegraphics[width=8.6cm]{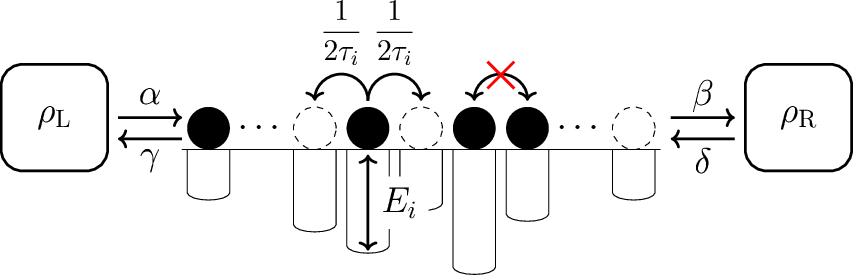}
    \caption{Schematic of the boundary-driven SEP on a quenched random energy landscape.
    Filled and open circles denote occupied and empty sites, respectively.}
    \label{fig: model}
\end{figure}

We consider the SEP on a quenched random energy landscape under open boundary conditions [see Fig.~\ref{fig: model}].
The energy landscape is quenched, that is, it remains fixed in time.
The system consists of a one-dimensional lattice with $L$ sites and the lattice constant is set to unity.
Hard-core exclusion restricts the occupation of each site to at most one particle.
At a bulk site $i=2,\ldots,L-1$, a particle attempts to hop to each neighboring site with rate $1/(2\tau_i)$, and the hop is accepted only when the target site is empty.
Here, $\tau_i$ denotes the mean waiting time for a hop attempt from site $i$.
It is determined by the depth $E_i$ of the energy trap at site $i$ through the Arrhenius law, $\tau_i=\tau_c\exp(E_i/T)$, where $T$ is the temperature and $\tau_c$ characterizes the microscopic time scale.
Each bulk site contains an energy trap with a random depth.
The trap depths $\{E_i\}$ are independent and identically distributed random variables drawn from the exponential distribution, $\phi(E)=T_g^{-1}\exp(-E/T_g)$, where $T_g$ is the glass temperature.
Through the Arrhenius law, the exponential distribution of the trap depths induces a power-law distribution of the mean waiting times,
\begin{equation}
    \psi_{\nu}(\tau)=\nu\tau_c^{\nu}\tau^{-1-\nu}
    \quad (\tau\geq \tau_c)
\end{equation}
with $\nu=T/T_g$~\cite{Bouchaud:1990aa}.
In the following, we focus on $\nu>1$, for which the mean waiting time is finite.

We next describe the boundary conditions.
The left and right boundaries are connected to particle reservoirs with densities $\rho_\mathrm{L}$ and $\rho_\mathrm{R}$, respectively.
To impose these reservoir densities, particles are injected into and ejected from the boundary sites with rates
\begin{equation}
    \begin{split}
        \alpha=\frac{\rho_{\mathrm{L}}}{2\tau_{\mathrm{r}}},\quad
        \gamma=\frac{1-\rho_{\mathrm{L}}}{2\tau_{\mathrm{r}}},\quad
        \delta=\frac{\rho_{\mathrm{R}}}{2\tau_{\mathrm{r}}},\quad
        \beta=\frac{1-\rho_{\mathrm{R}}}{2\tau_{\mathrm{r}}},
    \end{split}
\end{equation}
where $\tau_{\mathrm{r}}$ is the time scale of the exchange with the reservoirs.
At the left boundary, $\alpha$ and $\gamma$ are the injection and ejection rates, respectively, whereas at the right boundary, $\delta$ and $\beta$ are the corresponding rates.
Unlike the bulk sites, the boundary sites are not disordered.
To avoid introducing an additional source of left-right asymmetry at the boundaries and to isolate the effect of quenched disorder in the bulk, we set the mean waiting times at the two boundary sites to the same nonrandom value, $\tau_1=\tau_L=\tau_{\mathrm{s}}$.
A particle at a boundary site attempts to hop to its neighboring bulk site with rate $1/(2\tau_{\mathrm{s}})$.
When the reservoir densities are equal, $\rho_{\mathrm{L}}=\rho_{\mathrm{R}}=\rho$, the system is in equilibrium.
By contrast, when $\rho_{\mathrm{L}}-\rho_{\mathrm{R}}=\Delta\rho \neq 0$, the density difference drives the system out of equilibrium.
In this work, we parametrize the reservoir densities as
\begin{equation}
    \rho_{\mathrm{L}}=\rho+\frac{\Delta\rho}{2},\quad
    \rho_{\mathrm{R}}=\rho-\frac{\Delta\rho}{2},
\end{equation}
where $|\Delta\rho|\leq 2\min(\rho,1-\rho)$ and $\rho=(\rho_{\mathrm{L}}+\rho_{\mathrm{R}})/2$ is the mean reservoir density.

The numerical simulations are initialized with an empty lattice and performed using a continuous-time Monte Carlo method.
For each fixed disorder realization, we evolve the system for a sufficiently long relaxation time and begin measurements only after the system has reached a stationary state.

Our main observable is the steady-state current.
Let $Q_i(t)$ be the integrated current across the bond between sites $i$ and $i+1$ up to time $t$, where $i=1,\ldots,L-1$.
Jumps from $i$ to $i+1$ are counted as positive and jumps in the opposite direction are counted as negative. 
The instantaneous ensemble-averaged current is
\begin{equation}
    \begin{split}
        J_i(t)\equiv&\frac{d}{dt}\Braket{Q_i(t)}\\
        =&\dfrac{1}{2\tau_i}\Braket{\eta_i(1-\eta_{i+1})}-\dfrac{1}{2\tau_{i+1}}\Braket{\eta_{i+1}(1-\eta_i)},
    \end{split}
    \label{eq: bulk current}
\end{equation}
where $\eta_i\in\{0,1\}$ denotes the occupation number at site $i$, and $\Braket{\cdot}$ denotes an average over the stochastic dynamics for a fixed disorder realization.
In the steady state, current conservation ensures that the current is independent of the bond index $i$. 
We therefore define the steady-state current as
\begin{equation}
    J(\rho,\Delta \rho)\equiv\lim_{t\rightarrow\infty}\frac{d\Braket{Q_i(t)}}{dt}.
    \label{eq: definition of steady-state current}
\end{equation}
In the homogeneous environment, where $\tau_i=\tau$ for all sites, the steady-state current is given by~\cite{Derrida:2004aa}
\begin{equation}
    J(\rho,\Delta\rho)=\frac{\Delta\rho}{2\left[(L-1)\tau+2\tau_{\mathrm{r}}\right]}.
    \label{eq: current in homogeneous}
\end{equation}
Thus, in a homogeneous environment, the steady-state current is a linear function of the density difference and is independent of the mean reservoir density.

\section{Density response}\label{sec: density profile}
In this section, we demonstrate that the steady-state density profile exhibits a nonlinear response to the density difference.
We first derive the equilibrium distribution for the site-disordered SEP and then obtain the linear response of the steady-state probability using a Galerkin projection.
To extend the analysis beyond linear order, we adopt a closure that neglects connected nearest-neighbor correlations at second and higher orders.
The resulting approximate density profile, given by Eq.~\eqref{eq: nonlinear density profile}, depends on both the density difference $\Delta\rho$ and the mean reservoir density $\rho$.
We then compare the approximate expression with numerical simulations.

\subsection{Master equation}
We describe the dynamics using the master equation for the probability $P(\boldsymbol{\eta},t)$ of finding the system in configuration $\boldsymbol{\eta}=\{\eta_i\}_{i=1,\dots,L}$.
Because the nonequilibrium driving is introduced only through the boundary reservoirs, it is convenient to decompose the operator into the equilibrium part and the perturbation proportional to the density difference $\Delta\rho$.
The master equation is then written as
\begin{equation}
    \frac{\partial}{\partial t}P(\boldsymbol{\eta},t)
    =\left(\mathcal{L}_0+\Delta\rho\mathcal{L}_1\right)P(\boldsymbol{\eta},t),
\end{equation}
where the linear operators $\mathcal{L}_0$ and $\mathcal{L}_1$ are defined by
\begin{equation}
    \mathcal{L}_0f(\boldsymbol{\eta})
    \equiv\sum_{\boldsymbol{\eta}'}\left[W(\boldsymbol{\eta}',\boldsymbol{\eta})f(\boldsymbol{\eta}')
    -W(\boldsymbol{\eta},\boldsymbol{\eta}')f(\boldsymbol{\eta})\right]
\end{equation} 
and
\begin{equation}
    \begin{split}
        &\mathcal{L}_1f(\boldsymbol{\eta})\\
        \equiv&\frac{2\eta_1-1}{4\tau_{\mathrm{r}}}\left[f(\boldsymbol{\eta}^1)+f(\boldsymbol{\eta})\right]
        +\frac{1-2\eta_L}{4\tau_{\mathrm{r}}}\left[f(\boldsymbol{\eta}^L)+f(\boldsymbol{\eta})\right],
    \end{split}
    \label{eq: L1}
\end{equation}
respectively.
Here, $W(\boldsymbol{\eta}',\boldsymbol{\eta})$ denotes the transition rate from $\boldsymbol{\eta}'$ to $\boldsymbol{\eta}$ at $\Delta\rho=0$.
The configuration $\boldsymbol{\eta}^i$ is obtained from $\boldsymbol{\eta}$ by flipping the occupation number at site $i$, namely, $\eta_i^{i}=1-\eta_i$ and $\eta_j^{i}=\eta_j$ for $j\neq i$.
Thus, $\mathcal{L}_0$ describes the equilibrium dynamics, whereas $\Delta\rho\mathcal{L}_1$ represents the boundary-driving perturbation.
In the following, we derive the steady-state probability $P_{\mathrm{st}}(\boldsymbol{\eta})$, which satisfies $\partial P(\boldsymbol{\eta},t)/\partial t=0$.

\subsection{Equilibrium state}
We first determine the stationary probability $P_{\mathrm{eq}}(\boldsymbol{\eta})$ at equilibrium ($\Delta\rho=0$).
From the detailed balance relation, $W(\boldsymbol{\eta}',\boldsymbol{\eta})P_{\mathrm{eq}}(\boldsymbol{\eta}')=W(\boldsymbol{\eta},\boldsymbol{\eta}')P_{\mathrm{eq}}(\boldsymbol{\eta})$, we derive the probability $P_{\mathrm{eq}}(\boldsymbol{\eta})$:
\begin{equation}
    P_{\mathrm{eq}}(\boldsymbol{\eta})=\frac{1}{Z}\prod_{i=1}^L\left(1-\rho\right)^{1-\eta_i}\left(\frac{\tau_i}{\tau_{\mathrm{s}}}\rho\right)^{\eta_i},
\end{equation}
where $Z$ is the normalization constant.
From this distribution, the site density and the two-point correlation function at equilibrium are given by
\begin{equation}
    \Braket{\eta_i}_{\mathrm{eq}}=\frac{\dfrac{\tau_i}{\tau_{\mathrm{s}}}\rho}{1-\rho+\dfrac{\tau_i}{\tau_{\mathrm{s}}}\rho}\equiv \rho_i^{\mathrm{eq}}
    \label{eq: equilibrium site density}
\end{equation}
and
\begin{equation}
    \Braket{\eta_i\eta_j}_{\mathrm{eq}}
    =\frac{\dfrac{\tau_i\tau_j}{\tau_{\mathrm{s}}^2}\rho^2}{\left(1-\rho+\dfrac{\tau_i}{\tau_{\mathrm{s}}}\rho\right)\left(1-\rho+\dfrac{\tau_j}{\tau_{\mathrm{s}}}\rho\right)}
    =\rho_i^{\mathrm{eq}}\rho_j^{\mathrm{eq}}
\end{equation}
for $i\neq j$, respectively.
Here, $\Braket{\cdot}_{\mathrm{eq}}$ denotes the ensemble average with respect to $P_{\mathrm{eq}}(\boldsymbol{\eta})$.
Thus, the two-point correlation function factorizes into the product of the site densities, showing that the occupation variables at different sites are uncorrelated at equilibrium.

\subsection{Linear response}
We next derive the linear response of the steady-state probability $P_{\mathrm{st}}(\boldsymbol{\eta})$.
We expand $P_{\mathrm{st}}(\boldsymbol{\eta})$ perturbatively with respect to $\Delta\rho$ as
\begin{equation}
    P_{\mathrm{st}}(\boldsymbol{\eta})
    = P_{\mathrm{eq}}(\boldsymbol{\eta})\left[1+\Delta\rho \Phi_1(\boldsymbol{\eta})\right]
    +O((\Delta\rho)^2).
    \label{eq: expansion of steady-state probability}
\end{equation}
The term $P_{\mathrm{eq}}(\boldsymbol{\eta})\Phi_1(\boldsymbol{\eta})$ represents the linear correction to the steady-state probability.
The normalization of $P_{\mathrm{st}}(\boldsymbol{\eta})$ requires $\Braket{\Phi_1(\boldsymbol{\eta})}_{\mathrm{eq}}=0$.
Substituting Eq.~\eqref{eq: expansion of steady-state probability} into $\left(\mathcal{L}_0+\Delta\rho\mathcal{L}_1\right)P_{\mathrm{st}}(\boldsymbol{\eta})=0$ and collecting the terms proportional to $\Delta\rho$, we obtain
\begin{equation}
    \mathcal{L}_0\left[P_{\mathrm{eq}}(\boldsymbol{\eta})\Phi_1(\boldsymbol{\eta})\right]+\mathcal{L}_1P_{\mathrm{eq}}(\boldsymbol{\eta})
    =0.
    \label{eq: equation for Phi1}
\end{equation}
Using the detailed balance relation, $W(\boldsymbol{\eta}',\boldsymbol{\eta})P_{\mathrm{eq}}(\boldsymbol{\eta}')=W(\boldsymbol{\eta},\boldsymbol{\eta}')P_{\mathrm{eq}}(\boldsymbol{\eta})$, the first term in Eq.~\eqref{eq: equation for Phi1} can be written as
\begin{equation}
    \mathcal{L}_0\left[P_{\mathrm{eq}}(\boldsymbol{\eta})\Phi_1(\boldsymbol{\eta})\right]
    =P_{\mathrm{eq}}(\boldsymbol{\eta})\mathcal{L}_0^{\dagger}\Phi_1(\boldsymbol{\eta}),
\end{equation}
where $\mathcal{L}_0^{\dagger}$ is the adjoint operator of $\mathcal{L}_0$, defined by
\begin{equation}
    \mathcal{L}_0^{\dagger}f(\boldsymbol{\eta})=\sum_{\boldsymbol{\eta}'}W(\boldsymbol{\eta},\boldsymbol{\eta}')\left[f(\boldsymbol{\eta}')-f(\boldsymbol{\eta})\right].
\end{equation}
From Eq.~\eqref{eq: L1}, the second term in Eq.~\eqref{eq: equation for Phi1} is
\begin{equation}
    \begin{split}
        \mathcal{L}_1P_{\mathrm{eq}}(\boldsymbol{\eta})
        =&\frac{2\eta_1-1}{4\tau_{\mathrm{r}}}\left[\frac{P_{\mathrm{eq}}(\boldsymbol{\eta}^1)}{P_{\mathrm{eq}}(\boldsymbol{\eta})}+1\right]P_{\mathrm{eq}}(\boldsymbol{\eta})\\
        &+\frac{1-2\eta_L}{4\tau_{\mathrm{r}}}\left[\frac{P_{\mathrm{eq}}(\boldsymbol{\eta}^L)}{P_{\mathrm{eq}}(\boldsymbol{\eta})}+1\right]P_{\mathrm{eq}}(\boldsymbol{\eta}).
    \end{split}
    \label{eq: L1P_eq}
\end{equation}
The ratio $P_{\mathrm{eq}}(\boldsymbol{\eta}^i)/P_{\mathrm{eq}}(\boldsymbol{\eta})$ depends on the occupation number $\eta_i$ at site $i$ as
\begin{equation}
    \frac{P_{\mathrm{eq}}(\boldsymbol{\eta}^i)}{P_{\mathrm{eq}}(\boldsymbol{\eta})}
    =\frac{(1-\eta_i)\left(\dfrac{\tau_i}{\tau_{\mathrm{s}}}\rho\right)^2+\eta_i(1-\rho)^2}{\dfrac{\tau_i}{\tau_{\mathrm{s}}}\rho(1-\rho)}.
\end{equation}
Substituting this relation into Eq.~\eqref{eq: L1P_eq}, we obtain
\begin{equation}
    \mathcal{L}_1P_{\mathrm{eq}}(\boldsymbol{\eta})
    =P_{\mathrm{eq}}(\boldsymbol{\eta})\frac{\eta_1-\eta_L}{4\tau_{\mathrm{r}}\rho(1-\rho)}.
\end{equation}
Therefore, Eq.~\eqref{eq: equation for Phi1} can be rewritten as
\begin{equation}
    \mathcal{L}_0^{\dagger}\Phi_1(\boldsymbol{\eta})
    +\frac{\eta_1-\eta_L}{4\tau_{\mathrm{r}}\rho(1-\rho)}
    =0.
    \label{eq: equation for Phi1 2}
\end{equation}

We solve Eq.~\eqref{eq: equation for Phi1 2} approximately using the Galerkin approximation \cite{SCHUTTE1999146,10.1063/1.5063730}, in which $\Phi_1(\boldsymbol{\eta})$ is represented in a finite-dimensional subspace.
Because the equations for the one-site terms couple to nearest-neighbor two-site correlations, which also enter the expression for the current, we retain both types of basis functions.
We introduce the centered one-site variables $\xi_i\equiv \eta_i-\rho_i^{\mathrm{eq}}$ and approximate $\Phi_1(\boldsymbol{\eta})$ as
\begin{equation}
    \Phi_1(\boldsymbol{\eta})\simeq\sum_{i=1}^L a_i\xi_i+\sum_{i=1}^{L-1}b_i\xi_i\xi_{i+1}
    \equiv\Phi_1^{\mathrm{app}}(\boldsymbol{\eta}).
    \label{eq: expand Phi1}
\end{equation}
Substituting $\Phi_1^{\mathrm{app}}(\boldsymbol{\eta})$ into Eq.~\eqref{eq: equation for Phi1 2} leaves a nonzero residual.
We determine the coefficients $a_i$ and $b_i$ by requiring this residual to be orthogonal to all basis functions with respect to the equilibrium inner product.
These Galerkin conditions yield
\begin{equation}
    \Braket{\xi_i\left(\mathcal{L}_0^{\dagger}\Phi_1^{\mathrm{app}}(\boldsymbol{\eta})
    +\frac{\eta_1-\eta_L}{4\tau_{\mathrm{r}}\rho(1-\rho)}\right)}_{\mathrm{eq}}
    =0
    \label{eq: etai}
\end{equation}
for $i=1,\dots,L$, and
\begin{equation}
    \Braket{\xi_i\xi_{i+1}\left(\mathcal{L}_0^{\dagger}\Phi_1^{\mathrm{app}}(\boldsymbol{\eta})
    +\frac{\eta_1-\eta_L}{4\tau_{\mathrm{r}}\rho(1-\rho)}\right)}_{\mathrm{eq}}
    =0
    \label{eq: etai etai+1}
\end{equation}
for $i=1,\dots,L-1$.

Substituting Eq.~\eqref{eq: expand Phi1} into Eqs.~\eqref{eq: etai} and \eqref{eq: etai etai+1}, we obtain the following linear equations for $a_i$ and $b_i$:
\begin{widetext}
    \begin{equation}
        \begin{cases}
            \kappa_0a_1-\kappa_1(a_2-a_1)-\kappa_1d_1b_1=\dfrac{1}{4\tau_{\mathrm{r}}},\\[5pt]
            \kappa_{i-1}(a_i-a_{i-1})-\kappa_i(a_{i+1}-a_i)
            +\kappa_{i-1}d_{i-1}b_{i-1}-\kappa_id_ib_i
            =0 & (i=2,\dots,L-1),\\[5pt]
            \kappa_{L-1}(a_L-a_{L-1})+\kappa_La_L+\kappa_{L-1}d_{L-1}b_{L-1}=-\dfrac{1}{4\tau_{\mathrm{r}}},
        \end{cases}
        \label{eq: equation for a}
    \end{equation}
\end{widetext}
and
\begin{equation}
    \kappa_id_i(a_{i+1}-a_i)+\left(\kappa_{i-1}\chi_{i+1}+\kappa_id_i^2+\kappa_{i+1}\chi_i\right)b_i=0
    \label{eq: equation for b}
\end{equation}
for $i=1,\dots,L-1$.
Here, $\kappa_i$ denotes the equilibrium one-way hopping flux across the bond between sites $i$ and $i+1$,
\begin{equation}
    \kappa_i=\frac{\rho_i^{\mathrm{eq}}(1-\rho_{i+1}^{\mathrm{eq}})}{2\tau_i}=\frac{\rho_{i+1}^{\mathrm{eq}}(1-\rho_i^{\mathrm{eq}})}{2\tau_{i+1}}
\end{equation}
for $i=1,\dots,L-1$, with the boundary values $\kappa_0=\kappa_L=\rho(1-\rho)/(2\tau_{\mathrm{r}})$.
$\chi_i$ represents the local equilibrium density variance,
\begin{equation}
    \chi_i=\rho_i^{\mathrm{eq}}(1-\rho_i^{\mathrm{eq}}).
\end{equation}
$d_i=\rho_{i+1}^{\mathrm{eq}}-\rho_i^{\mathrm{eq}}$ denotes the difference in equilibrium density between neighboring sites.

Solving Eq.~\eqref{eq: equation for b} for $b_i$, we obtain
\begin{equation}
    b_i=-\frac{\kappa_id_i}{B_i}\left(a_{i+1}-a_i\right),
    \label{eq: equation for b 2}
\end{equation}
where $B_i=\kappa_{i-1}\chi_{i+1}+\kappa_id_i^2+\kappa_{i+1}\chi_i$.
Substituting Eq.~\eqref{eq: equation for b 2} into Eq.~\eqref{eq: equation for a}, we obtain a closed set of linear equations for $a_i$:
\begin{equation}
    \begin{cases}
        \kappa_0a_1-C_1(a_2-a_1)=\dfrac{1}{4\tau_{\mathrm{r}}},\\[5pt]
        C_{i-1}(a_i-a_{i-1})=C_i(a_{i+1}-a_i) & (i=2,\dots,L-1),\\[5pt]
        C_{L-1}(a_L-a_{L-1})+\kappa_La_L=-\dfrac{1}{4\tau_{\mathrm{r}}},
    \end{cases}
    \label{eq: eqaution for a 2}
\end{equation}
where $C_i=\kappa_i\left(1-\kappa_id_i^2/B_i\right)$.
Solving Eq.~\eqref{eq: eqaution for a 2}, we find
\begin{equation}
    a_i
    =\dfrac{\sum_{j=i}^{L-1}C_j^{-1}-\sum_{j=1}^{i-1}C_j^{-1}}{4\tau_{\mathrm{r}}\left(2+\kappa_0\sum_{j=1}^{L-1}C_j^{-1}\right)}
\end{equation}
Substituting this result into Eq.~\eqref{eq: equation for b 2}, we finally obtain
\begin{equation}
    b_i=\frac{d_i}{2\tau_{\mathrm{r}}\left(B_i-\kappa_id_i^2\right)\left(2+\kappa_0\sum_{j=1}^{L-1}C_j^{-1}\right)}.
\end{equation}

Using $a_i$ and $b_i$, we express the linear responses of the site density and the nearest-neighbor two-point correlation function as
\begin{equation}
    \Braket{\eta_i}
    \simeq\rho_i^{\mathrm{eq}}
    +a_i\chi_i\Delta\rho
    +O((\Delta\rho)^2)
    \label{eq: linear density}
\end{equation}
and
\begin{equation}
    \begin{split}
        \Braket{\eta_i\eta_{i+1}}
        \simeq&\rho_i^{\mathrm{eq}}\rho_{i+1}^{\mathrm{eq}}\\
        &+\left(a_i\chi_i\rho_{i+1}^{\mathrm{eq}}+a_{i+1}\rho_i^{\mathrm{eq}}\chi_{i+1}+b_i\chi_i\chi_{i+1}\right)\Delta\rho\\
        &+O((\Delta\rho)^2),
    \end{split}
    \label{eq: linear two correlation}
\end{equation}
respectively.
Here, $\Braket{\cdot}$ denotes a steady-state average over the stochastic dynamics for a fixed disorder realization.
In a homogeneous environment, the linear-response coefficients of the site densities are independent of $\rho$ \cite{Derrida:2002aa}.
In contrast, in a site-disordered system, these coefficients depend on $\rho$ through the equilibrium density profile.
As shown below, this dependence causes the linear current-response coefficient to vary with $\rho$.
The higher-order density responses additionally generate nonlinear current contributions, whose even-order components are responsible for rectification-like behavior.

\subsection{Nonlinear response}
We now approximate the nonlinear responses of the site density $\Braket{\eta_i}$.
We expand $\Braket{\eta_i}$ perturbatively with respect to $\Delta \rho$ as
\begin{equation}
    \Braket{\eta_i}
    \simeq\rho_i^{\mathrm{eq}}
    +a_i\chi_i\Delta\rho
    +\sum_{n=2}^{\infty}\left(\Delta\rho\right)^n\rho_i^{(n)}.
    \label{eq: site density MFA}
\end{equation}
To expand the Galerkin result beyond linear order, we adopt a hybrid closure for the nearest-neighbor correlation.
We retain the Galerkin contribution at linear order and approximate the second- and higher-order contributions in terms of the site-density responses:
\begin{equation}
    \begin{split}
        &\Braket{\eta_i\eta_{i+1}}\\
        \simeq&\rho_i^{\mathrm{eq}}\rho_{i+1}^{\mathrm{eq}}
        +\left(a_i\chi_i\rho_{i+1}^{\mathrm{eq}}+a_{i+1}\rho_i^{\mathrm{eq}}\chi_{i+1}+b_i\chi_i\chi_{i+1}\right)\Delta\rho\\
        &+\sum_{n=2}^{\infty}\left(\Delta\rho\right)^n\left[\rho_i^{\mathrm{eq}}\rho_{i+1}^{(n)}+\rho_i^{(n)}\rho_{i+1}^{\mathrm{eq}}+\sum_{m=1}^{n-1}\rho_i^{(m)}\rho_{i+1}^{(n-m)}\right],
    \end{split}
    \label{eq: two-point correlation MFA}
\end{equation}
where $\rho_i^{(1)}\equiv a_i\chi_i$.
Thus, the connected nearest-neighbor correlation is retained at linear order through the term $b_i\chi_i\chi_{i+1}\Delta\rho$, but its contributions at $O((\Delta\rho)^2)$ and higher are neglected.

\begin{figure*}[tbp]
    \centering
    \includegraphics[width=14.5cm]{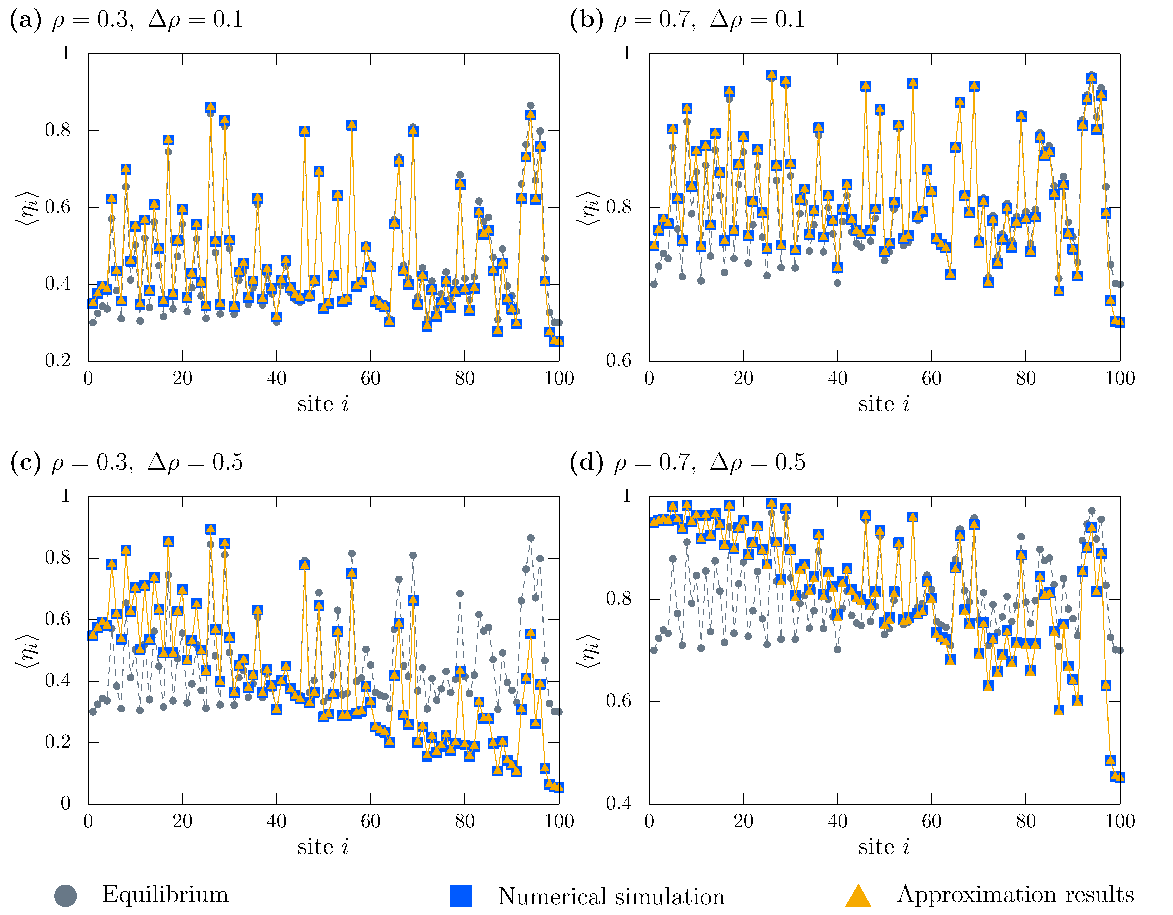}
    \caption{Steady-state density profiles for $L=100$, $\nu=1.5$, and $\tau_{\mathrm{r}}=\tau_{\mathrm{s}}=\tau_c=1$.
    Squares denote the numerical simulation results for the SEP on a quenched random energy landscape, triangles denote the approximation in Eq.~\eqref{eq: nonlinear density profile} truncated at $n_{\max}=10$, and circles denote the equilibrium density profile in Eq.~\eqref{eq: equilibrium site density}.
    The same disorder realization is used in all four panels.}
    \label{fig: Density profile}
\end{figure*}

To determine the higher-order responses $\rho_i^{(n)}$ within this closure, we impose the steady-state current conservation condition.
The bulk currents are given by Eq.~\eqref{eq: bulk current}.
We also define the boundary currents as
\begin{equation}
    J_0(t)\equiv\frac{d\Braket{Q_0(t)}}{dt}=\alpha(1-\Braket{\eta_1})-\gamma\Braket{\eta_1},
    \label{eq: boundary current}
\end{equation}
and
\begin{equation}
    J_L(t)\equiv\frac{d\Braket{Q_L(t)}}{dt}=\beta\Braket{\eta_L}-\delta(1-\Braket{\eta_L}).
\end{equation}
Here, $Q_0(t)$ and $Q_L(t)$ are the integrated current across the bonds between the boundary sites and their respective reservoirs.
In the steady state, the continuity equation gives
\begin{equation}
    J_{i-1}=J_i,
    \quad (i=1,\ldots,L),
\end{equation}
where $J_i=\lim_{t\rightarrow\infty}J_i(t)$.
Substituting Eqs.~\eqref{eq: site density MFA} and \eqref{eq: two-point correlation MFA} into the current-conservation condition, we obtain a recursive set of linear equations for $\rho_i^{(n)}$:
\begin{widetext}
    \begin{equation}
        \begin{cases}
            \kappa_0\dfrac{\rho_1^{(n)}}{\chi_1}
            -\kappa_1\left(\dfrac{\rho_2^{(n)}}{\chi_2}-\dfrac{\rho_1^{(n)}}{\chi_1}\right)
            =S_1^{(n)}\\[10pt]
            \kappa_{i-1}\left(\dfrac{\rho_i^{(n)}}{\chi_i}-\dfrac{\rho_{i-1}^{(n)}}{\chi_{i-1}}\right)
            -\kappa_i\left(\dfrac{\rho_{i+1}^{(n)}}{\chi_{i+1}}-\dfrac{\rho_i^{(n)}}{\chi_i}\right)
            =S_i^{(n)}-S_{i-1}^{(n)}
            & (i=2,\dots,L-1)\\[10pt]
            \kappa_{L-1}\left(\dfrac{\rho_L^{(n)}}{\chi_L}-\dfrac{\rho_{L-1}^{(n)}}{\chi_{L-1}}\right)
            +\kappa_L\dfrac{\rho_L^{(n)}}{\chi_L}
            =-S_{L-1}^{(n)}.
        \end{cases}
        \label{eq: equation for rhoin}
    \end{equation}
\end{widetext}
Here,
\begin{equation}
    S_i^{(n)}=\frac{1}{2}\left(\frac{1}{\tau_i}-\frac{1}{\tau_{i+1}}\right)\sum_{m=1}^{n-1}\rho_i^{(m)}\rho_{i+1}^{(n-m)}.
\end{equation}
Solving Eq.~\eqref{eq: equation for rhoin}, we obtain
\begin{equation}
    \rho_i^{(n)}
    =\chi_i\left[\frac{\sum_{j=1}^{L-1}\kappa_j^{-1}S_j^{(n)}}{\sum_{j=0}^{L}\kappa_j^{-1}}\sum_{j=0}^{i-1}\kappa_j^{-1}
    -\sum_{j=1}^{i-1}\frac{S_j^{(n)}}{\kappa_j}\right].
    \label{eq: solution for rho n}
\end{equation}
Substituting Eq.~\eqref{eq: solution for rho n} into Eq.~\eqref{eq: site density MFA}, we obtain the approximate nonlinear density profile as
\begin{equation}
    \begin{split}
        \Braket{\eta_i}
        \simeq&\rho_i^{\mathrm{eq}}
        +a_i\chi_i\Delta\rho\\
        &+\chi_i\sum_{n=2}^{\infty}\left(\Delta\rho\right)^n
        \left[\frac{\sum_{j=0}^{i-1}\kappa_j^{-1}}{\sum_{j=0}^{L}\kappa_j^{-1}}\sum_{j=1}^{L-1}\frac{S_j^{(n)}}{\kappa_j}
        -\sum_{j=1}^{i-1}\frac{S_j^{(n)}}{\kappa_j}\right].
    \end{split}
    \label{eq: nonlinear density profile}
\end{equation}
In the numerical comparisons below, we truncate the series in Eq.~\eqref{eq: nonlinear density profile} at $n_{\max}=10$; convergence with respect to this truncation is examined in Appendix~\ref{sec: truncation order}.

\subsection{Comparison with numerical results}
Here, we compare the approximate density profiles with the numerical results.
Figure~\ref{fig: Density profile} shows the density profiles for different values of the mean reservoir density $\rho$ and density difference $\Delta\rho$.
The approximation agrees well with the numerical results for the disorder realization shown, including the strongly heterogeneous regions where the equilibrium density profile exhibits large spatial variations.

Even for the same density difference $\Delta\rho$, the deviation from the equilibrium density profile depends strongly on the mean reservoir density $\rho$.
At the lower mean reservoir density, $\rho=0.3$, the site densities near the low-density reservoir, whose density is $\rho_{\mathrm{R}}=\rho-\Delta\rho/2$, show a large decrease from the equilibrium profile
[see Fig.~\ref{fig: Density profile}(c)].
In contrast, at the higher mean reservoir density, $\rho=0.7$, the corresponding deviation is much smaller
[see Fig.~\ref{fig: Density profile}(d)].
This comparison shows that the density response already depends on the mean reservoir density.
Its consequences for the linear and nonlinear current responses are examined in the next section.

\section{Current response}\label{sec: current}
In this section, we characterize the response of the steady-state current $J(\rho,\Delta\rho)$, defined in Eq.~\eqref{eq: definition of steady-state current}, to the reservoir density difference for a fixed disorder realization.
We distinguish two effects of site disorder that arise at different orders in the density difference.
First, we derive the linear current-response coefficient and show that it depends on the mean reservoir density.
We then relate this coefficient to the equilibrium current-fluctuation coefficient.
Second, we derive an approximate expression for the nonlinear current response and show that its even-order contributions break the antisymmetry of the current under reversal of the density difference, producing rectification-like behavior.
Finally, we show that spatial-reflection symmetry of the equilibrium density profile guarantees current antisymmetry, so broken spatial-reflection symmetry is a necessary condition for rectification-like behavior.
We then use the approximate second-order current response to show that, for continuously distributed site disorder, rectification-like behavior occurs arbitrarily close to equilibrium for almost every disorder realization.

\begin{figure*}[tbp]
    \centering
    \includegraphics[width=15cm]{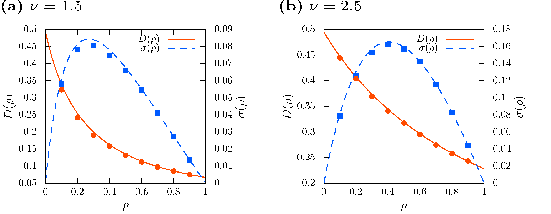}
    \caption{Dependence of the linear current-response coefficient $D(\rho)$ and the equilibrium current-fluctuation coefficient $\sigma(\rho)$ on the mean reservoir density $\rho$, for $L=100$ and $\tau_{\mathrm{r}}=\tau_{\mathrm{s}}=\tau_c=1$.
    Panels (a) and (b) show the results for $\nu=1.5$ and $\nu=2.5$, respectively.
    Circles denote the numerical estimates $D_{\mathrm{num}}(\rho)=L[J(\rho,0.1)-J(\rho,-0.1)]/0.2$, and the solid lines show the approximation in Eq.~\eqref{eq: response coefficient}.
    Squares denote the numerical results for $\sigma(\rho)$, and the dashed lines show Eq.~\eqref{eq: mobility}.
    The left and right vertical axes correspond to $D(\rho)$ and $\sigma(\rho)$, respectively.
    Within each panel, the same disorder realization is used for all values of $\rho$; different realizations are used in panels (a) and (b).}
    \label{fig: current fluctuation}
\end{figure*}

\subsection{Linear response}
We first consider the linear response of the steady-state current to the reservoir density difference.
When the density difference is small, we define the linear current-response coefficient $D(\rho)$ through
\begin{equation}
    \lim_{\Delta\rho\rightarrow0}\frac{J(\rho,\Delta\rho)}{\Delta\rho}=\frac{D(\rho)}{L}.
\end{equation}
At equilibrium, where $\Delta\rho=0$, the mean current vanishes, $\Braket{Q_i(t)}_{\mathrm{eq}}=0$.
We define the equilibrium current fluctuation coefficient $\sigma(\rho)$ by the long-time growth rate of the second moment of the integrated current,
\begin{equation}
    \lim_{t\rightarrow\infty}\frac{\Braket{Q_i^2(t)}_{\mathrm{eq}}}{t}=\frac{\sigma(\rho)}{L}.
\end{equation}
For each fixed disorder realization, the equilibrium dynamics
satisfies detailed balance. To relate the current-response
coefficient to the equilibrium current fluctuations, we express
the boundary driving in terms of the thermodynamic force
conjugate to particle transfer. Let
\begin{equation}
    \Delta\mu
    \equiv \mu(\rho_{\mathrm{L}})-\mu(\rho_{\mathrm{R}})
\end{equation}
denote the chemical-potential difference between the reservoirs.
The linear-response relation for the current then gives
\cite{Derrida:2007ab,SciPostPhysLectNotes.106}
\begin{equation}
    \lim_{\Delta\mu\rightarrow0}
    \frac{J}{\Delta\mu}
    =
    \frac{\sigma(\rho)}{2TL}.
    \label{eq: current response to chemical potential}
\end{equation}
This relation follows from detailed balance and does not require
spatially homogeneous hopping rates.

Using
$\rho_{\mathrm{L}}=\rho+\Delta\rho/2$ and
$\rho_{\mathrm{R}}=\rho-\Delta\rho/2$, the chemical-potential
difference is expanded as
\begin{equation}
    \Delta\mu
    =
    \frac{d\mu(\rho)}{d\rho}\Delta\rho
    +O((\Delta\rho)^2).
\end{equation}
Combining this expansion with the definition of $D(\rho)$ yields
\begin{equation}
    2D(\rho)
    =
    \frac{\sigma(\rho)}{T}
    \frac{d\mu(\rho)}{d\rho}.
    \label{eq: linear response relation for chemical potential}
\end{equation}
For the exclusion reservoirs considered here, the chemical potential is represented by $\mu(\rho)=T\log(\rho/(1-\rho))$.
Hence, the linear response relation is rewritten as
\begin{equation}
    2D(\rho)=\frac{\sigma(\rho)}{\rho(1-\rho)}.
    \label{eq: linear response relation}
\end{equation}
Thus, the linear-response relation in Eq.~\eqref{eq: linear response relation} holds for each fixed disorder realization, both in the homogeneous SEP and in the present site-disordered system.

We next evaluate the density dependence of $D(\rho)$ and $\sigma(\rho)$.
Substituting the linear density profile in Eq.~\eqref{eq: linear density} into Eq.~\eqref{eq: boundary current}, the response coefficient is obtained as 
\begin{equation}
    D(\rho)\simeq\frac{L}{4\tau_{\mathrm{r}}+\rho(1-\rho)\sum_{i=1}^{L-1}C_i^{-1}}.
    \label{eq: response coefficient}
\end{equation}
The response coefficient is independent of $\rho$ in a homogeneous environment~\cite{Derrida:2004aa}, whereas in the site-disordered environment it varies with $\rho$.
For the disorder realizations shown in Fig.~\ref{fig: current fluctuation}, $D(\rho)$ decreases monotonically with increasing $\rho$.
Using the linear response relation in Eq.~\eqref{eq: linear response relation}, $\sigma(\rho)$ is given by
\begin{equation}
    \sigma(\rho)\simeq\frac{2L\rho(1-\rho)}{4\tau_{\mathrm{r}}+\rho(1-\rho)\sum_{i=1}^{L-1}C_i^{-1}}.
    \label{eq: mobility}
\end{equation}
In a homogeneous environment, $\sigma(\rho)$ is symmetric under $\rho\rightarrow 1-\rho$~\cite{Derrida:2004aa}, whereas this symmetry is broken in the site-disordered environment [see Fig.~\ref{fig: current fluctuation}].
This asymmetry reflects the breaking of particle-hole symmetry by the site disorder.

\begin{figure*}[tbp]
    \centering
    \includegraphics[width=15cm]{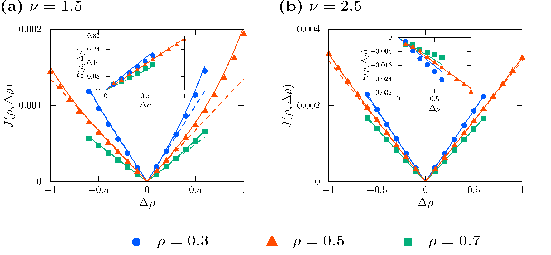}
    \caption{Magnitude of the steady-state current as a function of the density difference $\Delta\rho$, for $L=100$ and $\tau_{\mathrm{r}}=\tau_{\mathrm{s}}=\tau_c=1$.
    Panels (a) and (b) show the results for $\nu=1.5$ and $\nu=2.5$, respectively.
    Symbols denote the numerical simulation results.
    The solid lines show $|J(\rho,\Delta\rho)|$ obtained from Eq.~\eqref{eq: steady state current}, with the nonlinear series truncated at $n_{\max}=10$, and the dashed lines show its linear contribution.
    The insets show the logarithmic current ratio $R(\rho,\Delta\rho)$ defined in Eq.~\eqref{eq: logarithmic current ratio}.
    Within each panel, the same disorder realization is used for all values of $\rho$; different realizations are used in panels (a) and (b).}
    \label{fig: steady-state current}
\end{figure*}

\subsection{Nonlinear response}
We derive the steady-state current from the nonlinear density response obtained in the previous section.
Substituting the nonlinear density profile in Eq.~\eqref{eq: nonlinear density profile} into Eq.~\eqref{eq: boundary current}, we obtain
\begin{equation}
    \begin{split}
        J(\rho,\Delta\rho)
        \simeq&\frac{\Delta\rho}{4\tau_{\mathrm{r}}+\rho(1-\rho)\sum_{i=1}^{L-1}C_i^{-1}}\\
        &-\sum_{n=2}^{\infty}\left(\Delta\rho\right)^n\frac{\sum_{i=1}^{L-1}\kappa_i^{-1}S_i^{(n)}}{\sum_{i=0}^{L}\kappa_i^{-1}}.
    \end{split}
    \label{eq: steady state current}
\end{equation}
The first term gives the linear response of the current to the density difference, whereas the remaining terms contain the nonlinear corrections generated by the spatial heterogeneity of the waiting times.
In a homogeneous environment, $S_i^{(n)}=0$ for all $i$ because $\tau_i=\tau_{i+1}$.
Thus, the nonlinear corrections vanish within the present approximation, and the current is linear in $\Delta\rho$.
In contrast, in a site-disordered system, spatial variations in waiting times yield nonzero $S_i^{(n)}$, leading to a nonlinear current response.
Furthermore, since the coefficients $C_i$, $\kappa_i$, and $S_i^{(n)}$ depend on the mean reservoir density $\rho$, the steady-state current also depends on $\rho$.
In the numerical evaluation of Eq.~\eqref{eq: steady state current}, the nonlinear series is truncated at $n_{\max}=10$.
The truncation-order dependence is examined in Appendix A.

Equation~\eqref{eq: steady state current} also reveals how the asymmetry under reversal of the density difference arises.
Under $\Delta\rho\rightarrow -\Delta\rho$, the odd-order terms change sign, whereas the even-order terms remain unchanged.
Therefore, when odd- and even-order contributions coexist, the current is generally not antisymmetric,
\begin{equation}
    J(\rho,\Delta\rho)\neq -J(\rho,-\Delta\rho).
\end{equation}
Consequently, the current magnitudes for opposite density differences can differ.
Within the present approximation, the even-order coefficients are generated by the source terms $S_i^{(2m)}$, which combine the spatial variation of the hopping rates with the lower-order density responses.

We next examine whether these two predictions of Eq.~\eqref{eq: steady state current}---the $\rho$ dependence and the reversal asymmetry---are observed in the numerical simulations.
Figure~\ref{fig: steady-state current} shows that Eq.~\eqref{eq: steady state current} closely reproduces the simulated magnitude of the steady-state current as a function of the density difference $\Delta\rho$.
In contrast to the homogeneous SEP, the steady-state current is not simply proportional to $\Delta\rho$.
It also depends on the mean reservoir density $\rho$.
This $\rho$ dependence is a characteristic feature of the site-disordered system.

\subsection{Rectification-like behavior}
We now focus on the breaking of current antisymmetry under reversal of the density difference.
Figure~\ref{fig: steady-state current} shows that the magnitude of the steady-state current is not invariant under reversal of the density difference, $\Delta\rho\to-\Delta\rho$.
We refer to this asymmetry as rectification-like behavior and quantify it using the logarithmic current ratio
\begin{equation}
    R(\rho,\Delta\rho)
    =\log\left[\frac{J(\rho,\Delta\rho)}{-J(\rho,-\Delta\rho)}\right]\quad
    (\Delta\rho>0).
    \label{eq: logarithmic current ratio}
\end{equation}
This quantity compares the current magnitudes generated by opposite density differences.
When $R(\rho,\Delta\rho)=0$, the current magnitude is invariant under the reversal $\Delta\rho\to-\Delta\rho$.
By contrast, a nonzero value of $R(\rho,\Delta\rho)$ indicates rectification-like behavior.
The magnitude $|R(\rho,\Delta\rho)|$ measures the strength of the asymmetry.
Its sign specifies which direction of the density gradient produces the larger current magnitude: $R>0$ means that $|J(\rho,\Delta\rho)|>|J(\rho,-\Delta\rho)|$, whereas $R<0$ indicates the opposite inequality.

The insets of Figs.~\ref{fig: steady-state current}(a) and \ref{fig: steady-state current}(b) show that the logarithmic current ratio deviates from zero.
Its magnitude increases nearly monotonically with positive $\Delta\rho$ over the range shown.
Thus, although reversing the density difference reverses the direction of the current, it does not generally preserve the current magnitude.
The sign and magnitude of $R(\rho,\Delta\rho)$ depend on both the mean reservoir density $\rho$ and the disorder realization.

\subsection{Necessary condition for rectification-like behavior}
We next establish a necessary condition for rectification-like behavior by first identifying a sufficient condition for its absence.
For a disorder realization $\boldsymbol{\tau}=(\tau_1,\dots,\tau_L)$, let $\boldsymbol{\tau}^{\mathrm R}$ denote its spatially reflected realization, defined by $\tau_i^{\mathrm R}=\tau_{L+1-i}$.
Spatial reflection exchanges the two reservoirs and reverses the positive direction of the current.
Therefore,
\begin{equation}
    J_{\boldsymbol{\tau}^{\mathrm{R}}}(\rho,\Delta\rho)
    =-J_{\boldsymbol{\tau}}(\rho,-\Delta\rho).
    \label{eq: spatial reflection invariant}
\end{equation}
If the disorder realization $\boldsymbol{\tau}$ is invariant under spatial reflection, i.e., $\tau_i=\tau_{L+1-i}$ for all $i$ or $\boldsymbol{\tau}^{\mathrm R}=\boldsymbol{\tau}$, Eq.~\eqref{eq: spatial reflection invariant} gives
\begin{equation}
    J_{\boldsymbol{\tau}}(\rho,\Delta\rho)
    =-J_{\boldsymbol{\tau}}(\rho,-\Delta\rho).
\end{equation}
Consequently, the current magnitudes are identical for opposite density differences, and $R(\rho,\Delta\rho)=0$.
In the present model, the equilibrium site density is a strictly monotonic function of $\tau_i$.
Therefore, spatial-reflection symmetry of the disorder realization is equivalent to spatial-reflection symmetry of the equilibrium density profile:
\begin{equation}
    \rho_i^{\mathrm{eq}}=\rho_{L+1-i}^{\mathrm{eq}}\quad (i=1,\dots,L).
    \label{eq: reflection equilibrium density}
\end{equation}
It follows that spatial-reflection symmetry of the equilibrium density profile is a sufficient condition for the absence of rectification-like behavior.
By contraposition, broken spatial-reflection symmetry of the equilibrium density profile is therefore a necessary condition for rectification-like behavior.
This symmetry argument is exact and does not rely on the Galerkin projection or the higher-order closure used above.

We next consider how restrictive this necessary condition is for continuously distributed site disorder.
Since the waiting times in the bulk are independently drawn from a continuous distribution, the probability that $\tau_i=\tau_{L+1-i}$ for all $i$ is zero.
Thus, $\boldsymbol{\tau}\neq \boldsymbol{\tau}^{\mathrm{R}}$ almost surely, and hence the equilibrium density profile is almost surely reflection asymmetric.
This observation alone does not imply rectification-like behavior, because a reflection-asymmetric realization may, in principle, exhibit cancellations among nonlinear current contributions.

To examine whether such cancellations are typical, we focus on the second-order contribution to the nonlinear current response.
We expand the steady-state current as
\begin{equation}
    J_{\boldsymbol{\tau}}(\rho,\Delta\rho)
    =J_{\boldsymbol{\tau}}^{(1)}(\rho)\Delta\rho
    +J_{\boldsymbol{\tau}}^{(2)}(\rho)\left(\Delta\rho\right)^2
    +O((\Delta\rho)^3).
\end{equation}
This expansion gives
\begin{equation}
    J_{\boldsymbol{\tau}}(\rho,\Delta\rho)
    +J_{\boldsymbol{\tau}}(\rho,-\Delta\rho)
    =2J_{\boldsymbol{\tau}}^{(2)}(\rho)\left(\Delta\rho\right)^2
    +O((\Delta\rho)^4).
\end{equation}
Hence, if $J_{\boldsymbol{\tau}}^{(2)}(\rho)\neq0$, current antisymmetry is broken for arbitrarily small but nonzero $\Delta\rho$, resulting in rectification-like behavior.

Within the approximation in Eq.~\eqref{eq: steady state current}, $J_{\boldsymbol{\tau}}^{(2)}(\rho)$ is a rational function of the disorder variables $\boldsymbol{\tau}$.
The denominators appearing in this rational expression remain nonzero, and hence $J_{\boldsymbol{\tau}}^{(2)}(\rho)$ is a real analytic function.
Direct numerical evaluation confirms that, for each fixed value of $\rho$ examined, it is not identically zero as a function of $\boldsymbol{\tau}$.
Therefore, its zero set,
\begin{equation}
    \mathcal{Z}_2=\{\boldsymbol{\tau}:J_{\boldsymbol{\tau}}^{(2)}(\rho)=0\}
\end{equation}
has Lebesgue measure zero.
Because the disorder measure is absolutely continuous with respect to the Lebesgue measure, it follows that, for
each such fixed $\rho$,
\begin{equation}
    \Pr(J_{\boldsymbol{\tau}}^{(2)}=0)=0.
\end{equation}
Thus, $J_{\boldsymbol{\tau}}^{(2)}(\rho)\neq0$ almost surely for each value of $\rho$ examined.
Within the present approximation and for
these values of $\rho$, rectification-like behavior therefore occurs arbitrarily close to equilibrium for almost every disorder realization.

\begin{figure*}[tbp]
    \centering
    \includegraphics[width=15cm]{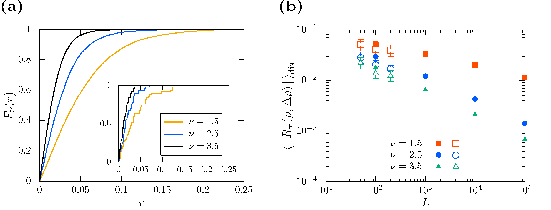}
    \caption{Cumulative distribution and disorder-averaged magnitude of the logarithmic current ratio for $\rho=0.5$, $\Delta\rho=0.5$, and $\tau_{\mathrm{r}}=\tau_{\mathrm{s}}=\tau_c=1$.
    The approximation in Eq.~\eqref{eq: steady state current}, truncated at $n_{\max}=10$, is compared with numerical simulations.
    (a) Cumulative distribution function $F_{\nu}(r)=\Pr(|R_{\boldsymbol{\tau}}(\rho,\Delta\rho)|\leq r)$ for $L=100$.
    The main panel shows the approximation obtained from $N_{\mathrm{dis}}=10^4$ independent disorder realizations for each $\nu$.
    The inset shows empirical cumulative distribution functions obtained from numerical simulations using $N_{\mathrm{dis}}=50$ independent disorder realizations for each $\nu$.
    (b) Disorder-averaged magnitude $\langle |R_{\boldsymbol{\tau}}(\rho,\Delta\rho)|\rangle_{\mathrm{dis}}$ as a function of the system size $L$ for $\nu=1.5$, $2.5$, and $3.5$.
    Filled symbols show the approximation using $N_{\mathrm{dis}}=10^4$ independent disorder realizations for each pair of $\nu$ and $L$.
    Open symbols show the numerical results using $N_{\mathrm{dis}}=50$, and the vertical error bars represent the 95\% bootstrap confidence intervals for $\langle |R_{\boldsymbol{\tau}}(\rho,\Delta\rho)|\rangle_{\mathrm{dis}}$ obtained by resampling the disorder realizations $10^5$ times.}
    \label{fig: current disorder}
\end{figure*}

\section{Disorder-ensemble symmetry and sample-to-sample fluctuations}\label{sec: disorder ensemble}
In this section, we examine the statistical properties of the steady-state current across disorder realizations.
We explicitly restore the disorder-realization label and write the steady-state current and the logarithmic current ratio as $J_{\boldsymbol{\tau}}(\rho,\Delta\rho)$ and $R_{\boldsymbol{\tau}}(\rho,\Delta\rho)$, respectively.
First, we establish the exact antisymmetry of the disorder-averaged current under reversal of the reservoir density difference.
We then derive the reflection symmetry of the sample-to-sample distribution of $R_{\boldsymbol{\tau}}(\rho,\Delta\rho)$.
Finally, we quantify the typical strength of the sample-specific rectification and examine its dependence on the system size.

\subsection{Ensemble symmetry and disorder-averaged current}
Let $P_{\mathrm{dis}}(\boldsymbol{\tau})$ denote the joint probability density of the disorder realization $\boldsymbol{\tau}$.
Because the disorder variables are independently drawn from the same distribution, the disorder ensemble is invariant under spatial reflection:
\begin{equation}
    P_{\mathrm{dis}}(\boldsymbol{\tau}^{\mathrm{R}})
    =P_{\mathrm{dis}}(\boldsymbol{\tau}).
    \label{eq: disorder probability distribution}
\end{equation}
Thus, each realization and its spatially reflected counterpart have the same statistical weight.
Using Eqs.~\eqref{eq: spatial reflection invariant} and \eqref{eq: disorder probability distribution}, and changing the integration variables from $\boldsymbol{\tau}$ to $\boldsymbol{\tau}^{\mathrm{R}}$, we obtain
\begin{equation}
    \Braket{J_{\boldsymbol{\tau}}(\rho,\Delta\rho)}_{\mathrm{dis}}=-\Braket{J_{\boldsymbol{\tau}}(\rho,-\Delta\rho)}_{\mathrm{dis}}.
    \label{eq:disorder-averaged-current-antisymmetry}
\end{equation}
Here, $\langle\cdot\rangle_{\mathrm{dis}}$ denotes the average over disorder realizations.
Equation~\eqref{eq:disorder-averaged-current-antisymmetry} shows that the disorder-averaged current is exactly antisymmetric under reversal of the density difference.
Thus, the disorder-averaged current exhibits no rectification-like behavior, even though individual disorder realizations can exhibit such behavior.

We next consider the sample-to-sample distribution of the logarithmic current ratio.
We define its probability density as
\begin{equation}
    P_{\nu}(r)=\Braket{\delta(r-R_{\boldsymbol{\tau}}(\rho,\Delta\rho))}_{\mathrm{dis}}.
\end{equation}
Using the current-reflection relation in Eq.~\eqref{eq: spatial reflection invariant}, together with the definition of the logarithmic current ratio, we obtain
\begin{equation}
    R_{\boldsymbol{\tau}}(\rho,\Delta\rho)
    =-R_{\boldsymbol{\tau}^{\mathrm{R}}}(\rho,\Delta\rho).
    \label{eq: R reflection antisymmetry}
\end{equation}
Combining this relation with the reflection invariance of the disorder ensemble in Eq.~\eqref{eq: disorder probability distribution}, and changing the integration variable from $\boldsymbol{\tau}$ to $\boldsymbol{\tau}^{\mathrm{R}}$, we obtain
\begin{equation}
    P_{\nu}(r)=P_{\nu}(-r).
    \label{eq: distribution of R symmetry}
\end{equation}
Thus, the distribution of $R_{\boldsymbol{\tau}}$ is exactly symmetric about $r=0$.
Consequently, all odd moments of $R_{\boldsymbol{\tau}}$ vanish.

We further consider the sample-specific current ratio
\begin{equation}
    \mathcal{A}_{\boldsymbol{\tau}}(\rho,\Delta\rho)
    =\frac{J_{\boldsymbol{\tau}}(\rho,\Delta\rho)}{-J_{\boldsymbol{\tau}}(\rho,-\Delta\rho)}
    =e^{R_{\boldsymbol{\tau}}(\rho,\Delta\rho)}
    \quad (\Delta\rho>0).
\end{equation}
The symmetry in Eq.~\eqref{eq: distribution of R symmetry} implies $\Braket{R_{\boldsymbol{\tau}}(\rho,\Delta\rho)}_{\mathrm{dis}}=0$.
Therefore, provided that $\Braket{\mathcal{A}_{\boldsymbol{\tau}}(\rho,\Delta\rho)}_{\mathrm{dis}}$ is finite, Jensen's inequality gives
\begin{equation}
    \begin{split}
        \Braket{\mathcal{A}_{\boldsymbol{\tau}}(\rho,\Delta\rho)}_{\mathrm{dis}}
        &=\Braket{e^{R_{\boldsymbol{\tau}}(\rho,\Delta\rho)}}_{\mathrm{dis}}\\
        &\geq e^{\Braket{R_{\boldsymbol{\tau}}(\rho,\Delta\rho)}_{\mathrm{dis}}}=1.
    \end{split}
\end{equation}
Because the exponential function is strictly convex, equality holds if and only if $R_{\boldsymbol{\tau}}=0$ for almost every
disorder realization.
Thus, the inequality is strict whenever sample-specific rectification occurs with nonzero probability.

This inequality does not imply a preferred transport direction.
Indeed, Eq.~\eqref{eq: R reflection antisymmetry} gives
\begin{equation}
    \mathcal{A}_{\boldsymbol{\tau}}=\mathcal{A}_{\boldsymbol{\tau}^{\mathrm{R}}}^{-1}
\end{equation}
Together with the reflection invariance of the disorder ensemble, this relation yields
\begin{equation}
    \Braket{\mathcal{A}_{\boldsymbol{\tau}}}_{\mathrm{dis}}
    =\Braket{\mathcal{A}_{\boldsymbol{\tau}}^{-1}}_{\mathrm{dis}}.
\end{equation}
Moreover, $\Braket{\mathcal{A}_{\boldsymbol{\tau}}}_{\mathrm{dis}}$ is the disorder average of a sample-specific current ratio and should not be confused with the ratio of disorder-averaged currents, which is equal to unity by Eq.~\eqref{eq:disorder-averaged-current-antisymmetry}.

For weak rectification, expansion of the exponential, together with the symmetry of $P_\nu(r)$, yields
\begin{equation}
    \Braket{\mathcal{A}_{\boldsymbol{\tau}}(\rho,\Delta\rho)}_{\mathrm{dis}}-1
    =\frac{1}{2}\Braket{R_{\boldsymbol{\tau}}^2}_{\mathrm{dis}}
    +\frac{1}{4!}\Braket{R_{\boldsymbol{\tau}}^4}_{\mathrm{dis}}+\cdots.
\end{equation}
Thus, to leading order, the excess of the disorder-averaged current ratio above unity is approximately the sample-to-sample fluctuations of $R_{\boldsymbol{\tau}}$.

\subsection{Distribution and finite-size behavior of rectification}
Although the disorder-averaged current is exactly antisymmetric under $\Delta\rho\rightarrow-\Delta\rho$, individual realizations can still exhibit rectification-like behavior, with realization-dependent signs and magnitudes.
Because the symmetry in Eq.~\eqref{eq: distribution of R symmetry} implies $\Braket{R_{\boldsymbol{\tau}}(\rho,\Delta\rho)}_{\mathrm{dis}}=0$, the signed disorder average does not characterize the typical strength of the sample-specific rectification.
We therefore use the disorder-averaged absolute value $\Braket{|R_{\boldsymbol{\tau}}(\rho,\Delta\rho)|}_{\mathrm{dis}}$.

Figure~\ref{fig: current disorder}(a) shows the cumulative distribution function (CDF) of $|R_{\boldsymbol{\tau}}|$, defined as $F_{\nu}(r)=\Pr(|R_{\boldsymbol{\tau}}|\leq r)$.
As $\nu$ decreases, the CDF shifts toward larger values of $r$, demonstrating that stronger disorder enhances the typical magnitude of the sample-specific rectification-like response.
The empirical CDFs obtained from numerical simulations exhibit the same ordering as the approximation results.
The additional results in Figs.~\ref{fig: SEP current disorder appendix}(a) and \ref{fig: SEP current disorder appendix}(b) show that, over the parameter ranges examined, the distribution also shifts toward smaller $r$ when $\Delta\rho$ is decreased at fixed $\rho$ or when $\rho$ is increased at fixed $\Delta\rho$.

Figure~\ref{fig: current disorder}(b) shows the dependence of $\Braket{|R_{\boldsymbol{\tau}}(\rho,\Delta\rho)|}_{\mathrm{dis}}$ on the system size $L$.
For the fixed reservoir parameters $\rho=0.5$ and $\Delta\rho=0.5$, both the numerical simulations and the approximation show that $\Braket{|R_{\boldsymbol{\tau}}(\rho,\Delta\rho)|}_{\mathrm{dis}}$ is larger for smaller $\nu$, indicating that stronger disorder enhances the sample-specific rectification-like response.
Furthermore, $\Braket{|R_{\boldsymbol{\tau}}(\rho,\Delta\rho)|}_{\mathrm{dis}}$ decreases with increasing $L$ for all values of $\nu$ examined.
A similar decrease is observed for other values of $\rho$ and $\Delta\rho$, as shown in Figs.~\ref{fig: SEP current disorder appendix}(c) and \ref{fig: SEP current disorder appendix}(d).
These results indicate that sample-specific rectification becomes weaker in larger systems over the range examined.
However, the available finite-size data do not determine its asymptotic behavior as $L\rightarrow\infty$.

\begin{table*}[tbp]
\caption{
Comparison of the current responses under density-gradient and external-field driving.
The density-gradient results for site disorder are obtained in the present work, whereas those for bond disorder are taken from Ref.~\cite{PhysRevLett.80.85}.
The external-field results are taken from Ref.~\cite{21vm-n8gp}.
}
\label{tab: comparison driving}
\begin{ruledtabular}
\begin{tabular}{llcc}
Property & Disorder & Density-gradient driving & External-field driving \\
\hline
Driving mechanism
& -- & Boundary reservoirs & Bulk hopping bias \\
Equilibrium density profile (in the absence of driving) & site & nonuniform & nonuniform\\
& bond & uniform & uniform\\
Linear-response coefficient
& site & $\rho$-dependent & Particle-hole asymmetric \\
& bond & $\rho$-independent & Particle-hole symmetric \\
Rectification-like behavior
& site & present & present \\
& bond & absent & absent \\
\end{tabular}
\end{ruledtabular}
\end{table*}

\section{Discussion}\label{sec: discussion}
We compare the present findings for density-gradient driving with our previous results for external-field driving~\cite{21vm-n8gp}.
The main similarities and differences between the two driving protocols are summarized in Table~\ref{tab: comparison driving}.
Although the equilibrium ensembles differ owing to the different boundary conditions, site disorder generates a spatially nonuniform equilibrium density profile in both systems.
The two driving protocols act on the dynamics differently, yet site disorder produces a reversal-asymmetric current response in both cases.
This correspondence suggests that the spatial structure of the equilibrium density profile, particularly its spatial-reflection asymmetry, plays a key role in the common rectification-like response, rather than the specific form of driving.

The result in Sec. IV D also provides a structural basis for understanding the different current responses of site- and bond-disordered systems.
The following discussion concerns the current response for a fixed disorder realization, rather than the disorder-averaged current.
In homogeneous and bond-disordered exclusion processes, the equilibrium density profile is spatially uniform and hence invariant under spatial reflection.
In both cases, the current is antisymmetric under reversal of the drive, as shown previously for boundary-driven and external-field-driven systems~\cite{Derrida:2004aa,21vm-n8gp,PhysRevLett.80.85,Derrida:1997aa}.
For a fixed realization of site disorder, the equilibrium density profile is generally nonuniform, but current antisymmetry is restored when the equilibrium profile remains invariant under spatial reflection.
Taken together, these findings show that, within the models considered here, spatial-reflection symmetry of the equilibrium density profile provides a sufficient criterion for the absence of rectification-like behavior.

Correspondingly, broken spatial-reflection symmetry of the equilibrium density profile is a necessary condition for rectification-like behavior at the sample level.
This condition is common to the boundary-driven system studied here and the external-field-driven system studied in Ref.~\cite{21vm-n8gp}.
The exact symmetry argument does not establish whether this condition is sufficient.
In principle, cancellations among nonlinear current contributions could restore current antisymmetry.
Thus, an asymmetric equilibrium density profile permits but does not guarantee rectification-like behavior.

Rectification in nanopores and nanochannels is often associated with spatially asymmetric geometries, surface charges, or internal energy landscapes~\cite{Bhattacharya:2011aa,Manara:2015aa,Zhou:2020aa}.
Our comparison between site and bond disorder shows, however, that spatial asymmetry of the microscopic disorder energy landscape alone does not necessarily generate a rectification-like response.
Both types of disorder can break spatial-reflection symmetry of the microscopic environment in an individual realization, but only site disorder produces a spatially nonuniform equilibrium density profile.
This distinction between microscopic environmental asymmetry and equilibrium-profile asymmetry may help clarify rectification mechanisms in more complex transport systems, including nanopores and nanochannels.

The present analytical approximation is restricted to the regime $\nu>1$.
Preliminary comparisons indicate that, for $\nu<1$, the approximation does not accurately reproduce the numerical results even for individual disorder realizations.
Clarifying the origin of this discrepancy and extending the approximation to this regime remain topics for future work.

\section{Conclusion}\label{sec: conclusion}
We have studied the current response of a boundary-driven symmetric exclusion process with quenched site disorder.
Combining a Galerkin approximation for the linear density response with a mean-field closure at higher orders, we found that site disorder produces a mean-density-dependent linear current response and a nonlinear reversal asymmetry.
Through the linear-response relation, the former breaks the $\rho\to1-\rho$ symmetry of the equilibrium current-fluctuation coefficient, while the latter gives rise to sample-specific rectification-like behavior.

At the sample level, we further showed that spatial-reflection symmetry of the equilibrium density profile is a sufficient condition to rule out rectification-like behavior.
Broken spatial-reflection symmetry of this profile is therefore a necessary condition for such behavior.
Within the present approximation, the second-order current coefficient is nonzero for almost every realization of continuously distributed site disorder, implying that, although absent in strict linear response, rectification-like behavior generically emerges at quadratic order arbitrarily close to equilibrium.
At the ensemble level, the disorder-averaged current remains exactly antisymmetric because the disorder distribution is invariant under spatial reflection.
Thus, the rectification-like response is sample-specific and cancels in the disorder-averaged current.

Finally, the disorder-averaged magnitude of the logarithmic current ratio decreases with increasing system size over the range examined.
This finite-size trend suggests that sample-specific rectification becomes weaker in larger systems, although the asymptotic behavior as $L\rightarrow\infty$ remains unresolved.
These results distinguish sample-specific reversal asymmetry from ensemble-level transport and show how spatial heterogeneity can generate rectification-like behavior without an explicitly imposed directional bias.

\begin{acknowledgments}
    I.S. was supported by JST SPRING Grant No. JPMJSP2151 and JSPS KAKENHI Grant No. JP26KJ2004.
\end{acknowledgments}

\appendix

\section{Truncation-order dependence}\label{sec: truncation order}
To examine the convergence of the nonlinear expansion, we denote the current evaluated by truncating Eq.~\eqref{eq: steady state current} at order $N$ by $J^{[N]}(\rho,\Delta\rho)$.
We define the relative difference between two even truncation orders $m>n$ as
\begin{equation}
    \varepsilon_J^{[m,n]}
    =\max_{s=\pm1}\frac{\left|J^{[m]}(\rho,s\Delta\rho)-J^{[n]}(\rho,s\Delta\rho)\right|}{\left|J^{[m]}(\rho,s\Delta\rho)\right|},
\end{equation}
where $s=\pm1$ accounts for both signs of the density difference, and $J^{[m]}$ is used as the higher-order reference value.
For the disorder realization used in Fig.~\ref{fig: steady-state current}(a), $\varepsilon_J^{[20,10]}=1.43\times 10^{-3}$ at $\rho=0.5$ and $|\Delta\rho|=1$.
This small difference indicates close agreement between the results obtained with $n_{\max}=10$ and $20$ for this realization and parameter set.
However, at $|\Delta\rho|=1$, the truncation order required for convergence can depend strongly on the disorder realization, and some realizations require orders higher than $10$.

We next assess the truncation-order dependence under the conditions used in the system-size analysis in Fig.~\ref{fig: current disorder}.
We focus on $\nu=1.5$, corresponding to the strongest disorder considered there, with $\rho=0.5$ and $|\Delta\rho|=0.5$.
As shown in Table~\ref{tab: truncation current}, the disorder average and the 95th percentile of $\varepsilon_J^{[20,10]}$ are at most $3.5\times10^{-6}$ and $1.4\times10^{-5}$, respectively.
The largest value observed across the validation samples is $5.7\times10^{-4}$.
These results show that the currents used in Fig.~\ref{fig: current disorder} are insensitive to increasing the truncation order from $n_{\max}=10$ to $20$ under the conditions examined.

We additionally examine the truncation accuracy for $R(\rho,\Delta\rho)$ under the conditions used in the system-size analysis in Fig.~\ref{fig: current disorder}.
For two even truncation orders $m>n$, we define the absolute difference as
\begin{equation}
    \Delta R^{[m,n]}
    =\left|R^{[m]}(\rho,\Delta\rho)-R^{[n]}(\rho,\Delta\rho)\right|.
\end{equation}
We focus on $\nu=1.5$, corresponding to the strongest disorder considered there, with $\rho=0.5$ and $\Delta\rho=0.5$.
As shown in Table~\ref{tab: truncation R}, the disorder average and the 95th percentile of $\Delta R^{[20,10]}$ are at most $1.3\times10^{-6}$ and $6.0\times10^{-6}$, respectively.

\begin{table}[tbp]
\caption{
Relative difference between truncation orders for $\nu=1.5$, $\rho=0.5$, and $|\Delta\rho|=0.5$.
For each system size, the relative differences were evaluated over $N_{\mathrm{dis}}=10^4$ independently generated disorder realizations.
The validation ensemble is independent of that used in Fig.~\ref{fig: current disorder}.
For each realization, the currents with $n_{\max}=10$ and $20$ were evaluated using the same disorder configuration.
Here, $P_{95}$ denotes the 95th percentile, and the maximum is the largest value observed among the $N_{\mathrm{dis}}$ realizations.
}
\label{tab: truncation current}
\begin{ruledtabular}
\begin{tabular}{cccc}
$L$
&
$\langle\varepsilon_J^{[20,10]}\rangle_{\mathrm{dis}}$
&
$P_{95}(\varepsilon_J^{[20,10]})$
&
$\max(\varepsilon_J^{[20,10]})$
\\
\hline
$10^{2}$ & $3.5\times10^{-6}$ & $1.4\times10^{-5}$
         & $5.3\times10^{-4}$ \\
$10^{3}$ & $2.8\times10^{-6}$ & $8.3\times10^{-6}$
         & $5.7\times10^{-4}$ \\
$10^{4}$ & $1.9\times10^{-6}$ & $3.9\times10^{-6}$ & $4.2\times10^{-4}$\\
$10^{5}$ & $1.5\times10^{-6}$ & $2.3\times10^{-6}$ & $9.9\times10^{-5}$\\
\end{tabular}
\end{ruledtabular}
\end{table}

\begin{table}[tbp]
\caption{
Absolute difference in $R(\rho,\Delta\rho)$ between truncation orders for $\nu=1.5$, $\rho=0.5$, and $\Delta\rho=0.5$.
For each system size, the absolute differences were evaluated over $N_{\mathrm{dis}}=10^4$ independently generated disorder realizations.
The validation ensemble is independent of that used in Fig.~\ref{fig: current disorder}.
For each realization, the currents with $n_{\max}=10$ and $20$ were evaluated using the same disorder configuration.
Here, $P_{95}$ denotes the 95th percentile.
}
\label{tab: truncation R}
\begin{ruledtabular}
\begin{tabular}{ccc}
$L$
&
$\langle\Delta R^{[20,10]}\rangle_{\mathrm{dis}}$
&
$P_{95}(\Delta R^{[20,10]})$
\\
\hline
$10^{2}$ & $1.3\times10^{-6}$ & $6.0\times10^{-6}$\\
$10^{3}$ & $9.2\times10^{-7}$ & $3.2\times10^{-6}$\\
$10^{4}$ & $4.9\times10^{-7}$ & $1.4\times10^{-6}$\\
$10^{5}$ & $2.5\times10^{-7}$ & $5.8\times10^{-7}$\\
\end{tabular}
\end{ruledtabular}
\end{table}

\begin{figure*}[tbp]
    \centering
    \includegraphics[width=15cm]{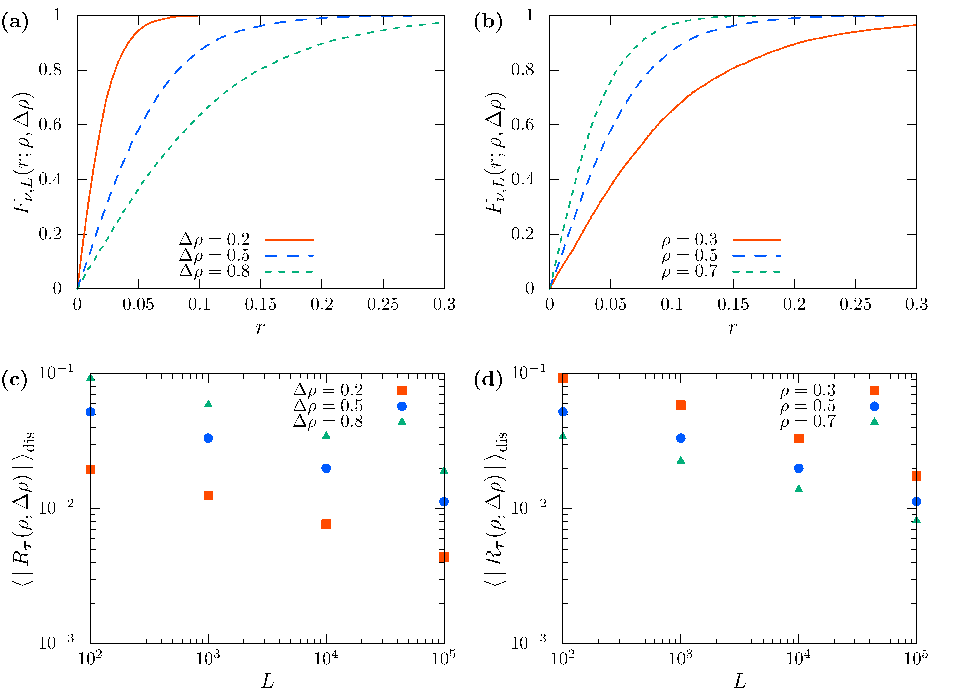}
    \caption{Dependence of the cumulative distribution and disorder-averaged magnitude of the logarithmic current ratio on the reservoir parameters for $\nu=1.5$ and $\tau_{\mathrm{r}}=\tau_{\mathrm{s}}=\tau_c=1$.
    All results are obtained from the approximate steady-state current in Eq.~\eqref{eq: steady state current}.
    The nonlinear series is truncated at $n_{\max}=20$ for $(\rho,\Delta\rho)=(0.5,0.8)$ and $(0.3,0.5)$, and at $n_{\max}=10$ for all other parameter sets.
    (a) Cumulative distribution function $F_{\nu,L}(r;\rho,\Delta\rho)=\Pr_{\mathrm{dis}}(|R_{\boldsymbol{\tau}}(\rho,\Delta\rho)|\leq r)$ for $L=100$ and fixed $\rho=0.5$, with the indicated values of $\Delta\rho$.
    (b) The same quantity for $L=100$ and fixed $\Delta\rho=0.5$, with the indicated values of $\rho$.
    The curves in (a) and (b) are obtained using $N_{\mathrm{dis}}=10^4$ independent disorder realizations for each parameter set.
    (c) Disorder-averaged magnitude of the logarithmic current ratio, $\langle |R_{\boldsymbol{\tau}}(\rho,\Delta\rho)|\rangle_{\mathrm{dis}}$, as a function of the system size $L$ for fixed $\rho=0.5$ and the indicated values of $\Delta\rho$.
    (d) The same quantity for fixed $\Delta\rho=0.5$ and the indicated values of $\rho$.
    The symbols in (c) and (d) are obtained using $N_{\mathrm{dis}}=10^4 $ independent disorder realizations for each parameter set and system size.}
    \label{fig: SEP current disorder appendix}
\end{figure*}

\section{Dependence on the reservoir parameters}
In Sec.~\ref{sec: disorder ensemble} B, we examined the finite-size dependence of sample-specific rectification for fixed reservoir parameters, $\rho=0.5$ and $\Delta\rho=0.5$.
Here, we examine whether the observed behavior persists for other values of the mean reservoir density and the density difference. We focus on $\nu=1.5$, which corresponds to the strongest disorder considered in Fig.~\ref{fig: current disorder}, and evaluate the logarithmic current ratio using the approximate steady-state current in Eq.~\eqref{eq: steady state current}.

Figures~\ref{fig: SEP current disorder appendix}(a) and \ref{fig: SEP current disorder appendix}(b) show the cumulative distribution function $F_{\nu,L}(r;\rho,\Delta\rho)=\Pr(|R_{\boldsymbol{\tau}}(\rho,\Delta\rho)|\leq r)$ for different reservoir parameters.
At fixed $\rho=0.5$, decreasing $\Delta\rho$ shifts the distribution of $|R_{\boldsymbol{\tau}}|$ toward smaller values, as shown in Fig.~\ref{fig: SEP current disorder appendix}(a).
Thus, a weaker density difference produces a weaker sample-specific rectification-like response.
At fixed $\Delta\rho=0.5$, increasing $\rho$ from $0.3$ to $0.7$ also shifts the distribution toward smaller values of $|R_{\boldsymbol{\tau}}|$ [Fig.~\ref{fig: SEP current disorder appendix}(b)].
These results show that both the density difference and the mean reservoir density affect both the nonlinear current and the sample-to-sample distribution of its reversal asymmetry.

Figures~\ref{fig: SEP current disorder appendix}(c) and \ref{fig: SEP current disorder appendix}(d) show the corresponding disorder-averaged magnitude $\langle |R_{\boldsymbol{\tau}}(\rho,\Delta\rho)|\rangle_{\mathrm{dis}}$ as a function of the system size.
For fixed $\rho$, its magnitude is larger for larger $\Delta\rho$, whereas for fixed $\Delta\rho$, it is larger for smaller $\rho$ over the range examined.
Nevertheless, $\langle |R_{\boldsymbol{\tau}}(\rho,\Delta\rho)|\rangle_{\mathrm{dis}}$ decreases with increasing $L$ for every parameter set considered.
The finite-size weakening of sample-specific rectification observed in Fig.~\ref{fig: current disorder} is therefore not restricted to the particular choice $\rho=0.5$ and $\Delta\rho=0.5$.
The present results, however, do not determine the asymptotic scaling with $L$ or establish whether the rectification measure vanishes in the thermodynamic limit.


\bibliography{SEP_open}

\end{document}